\documentclass[twocolumn]{aastex61}
\usepackage{amssymb}
\usepackage{amsmath}
\usepackage{graphicx}
\usepackage{CJK}
\newcommand{\bdv}[1]{\mbox{\boldmath$#1$}}

\def\e{{\rm E}}
\def\s{{\rm S}}
\def\lens{{\rm L}}
\def\b{{\rm B}}
\def\au{{\rm AU}}
\def\piee{{\pi_{\rm E,E}}}
\def\pien{{\pi_{\rm E,N}}}
\def\pie{{\pi_{\rm E}}}

\def\sat{{\rm sat}}
\def\jup{{\rm J}}

\shorttitle{Two-satellite Event MOA-2016-BLG-290}
\shortauthors{Zhu et al.}

\begin{document}
\begin{CJK*}{UTF8}{gbsn}
\title{An Isolated Microlens Observed from \emph{K2}, \emph{Spitzer} and Earth}

\author{Wei~Zhu (祝伟)}
\affil{Department of Astronomy, Ohio State University, 140 W. 18th Ave., Columbus, OH  43210, USA}
\affil{Canadian Institute for Theoretical Astrophysics, 60 St George Street, University of Toronto, Toronto, ON M5S 3H8, Canada}

\author{A.~Udalski}
\affil{Warsaw University Observatory, Al. Ujazdowskie 4, 00-478 Warszawa, Poland}

\author{C.~X.~Huang}
\affil{Department of Physics and Kavli Institute for Astrophysics and Space Research, Massachusetts Institute of Technology, Cambridge, MA 02139, USA}

\author{S.~Calchi~Novati}
\affil{IPAC, Mail Code 100-22, Caltech, 1200 E. California Blvd., Pasadena, CA 91125}
\affil{Dipartimento di Fisica ``E. R. Caianiello'', Universit\`a di Salerno, Via Giovanni Paolo II, 84084 Fisciano (SA), Italy}

\author{T.~Sumi}
\affil{Department of Earth and Space Science, Graduate School of Science, Osaka University, 1-1 Machikaneyama, Toyonake, Osaka 560-0043, Japan}

\nocollaboration
\correspondingauthor{Wei Zhu}
\email{zhu.908@osu.edu}
\author{R.~Poleski}
\affil{Department of Astronomy, Ohio State University, 140 W. 18th Ave., Columbus, OH 43210, USA}
\affil{Warsaw University Observatory, Al. Ujazdowskie 4, 00-478 Warszawa, Poland}
\author{J.~Skowron}
\affil{Warsaw University Observatory, Al. Ujazdowskie 4, 00-478 Warszawa, Poland}
\author{P.~Mr{\'o}z}
\affil{Warsaw University Observatory, Al. Ujazdowskie 4, 00-478 Warszawa, Poland}
\author{M.~K.~Szyma{\'n}ski}
\affil{Warsaw University Observatory, Al. Ujazdowskie 4, 00-478 Warszawa, Poland}
\author{I.~Soszy{\'n}ski}
\affil{Warsaw University Observatory, Al. Ujazdowskie 4, 00-478 Warszawa, Poland}
\author{P.~Pietrukowicz}
\affil{Warsaw University Observatory, Al. Ujazdowskie 4, 00-478 Warszawa, Poland}
\author{S.~Koz\l{}owski}
\affil{Warsaw University Observatory, Al. Ujazdowskie 4, 00-478 Warszawa, Poland}
\author{K.~Ulaczyk}
\affil{Warsaw University Observatory, Al. Ujazdowskie 4, 00-478 Warszawa, Poland}
\affil{Department of Physics, University of Warwick, Gibbert Hill Road, Coventry, CV4 7AL, UK}
\author{M.~Pawlak}
\affil{Warsaw University Observatory, Al. Ujazdowskie 4, 00-478 Warszawa, Poland}
\collaboration{(OGLE Collaboration)} \\
\author{C.~Beichman}
\affil{NASA Exoplanet Science Institute, MS 100-22, California Institute of Technology, Pasadena, CA 91125, USA}
\author{G.~Bryden}
\affil{Jet Propulsion Laboratory, California Institute of Technology, 4800 Oak Grove Drive, Pasadena, CA 91109, USA}
\author{S.~Carey}
\affil{Spitzer Science Center, MS 220-6, California Institute of Technology, Pasadena, CA, US}
\author{B.~S.~Gaudi}
\affil{Department of Astronomy, Ohio State University, 140 W. 18th Ave., Columbus, OH 43210, USA}
\author{A.~Gould}
\affil{Department of Astronomy, Ohio State University, 140 W. 18th Ave., Columbus, OH  43210, USA}
\affil{Korea Astronomy and Space Science Institute, 776 Daedeokdae-ro, Yuseong-Gu, Daejeon 34055, Korea}
\affil{Max-Planck-Institute for Astronomy, K\"onigstuhl 17, 69117 Heidelberg, Germany}
\author{C.~B.~Henderson}
\affil{Jet Propulsion Laboratory, California Institute of Technology, 4800 Oak Grove Drive, Pasadena, CA 91109, USA}
\author{Y.~Shvartzvald}
\affil{Jet Propulsion Laboratory, California Institute of Technology, 4800 Oak Grove Drive, Pasadena, CA 91109, USA}
\affil{NASA Postdoctoral Program Fellow}
\author{J.~C.~Yee}
\affil{Smithsonian Astrophysical Observatory, 60 Garden St., Cambridge, MA 02138, USA}
\collaboration{(Spitzer Team)}
\author{I.~A.~Bond}
\affil{Institute for Natural and Mathematical Sciences, Massey University, Private Bag 102904 North Shore Mail Centre, Auckland 0745, New Zealand}
\author{D.~P.~Bennett}
\affil{Laboratory for Exoplanets and Stellar Astrophysics, NASA/Goddard Space Flight Center, Greenbelt, MD 20771, USA}
\author{D.~Suzuki}
\affil{Institute of Space and Astronautical Science, Japan Aerospace Exploration Agency, 3-1-1 Yoshinodai, Chuo, Sagamihara, Kanagawa 252-5210, Japan}
\author{N.~J.~Rattenbury}
\affil{Department of Physics, University of Auckland, Private Bag 92019, Auckland, New Zealand}
\author{N.~Koshimoto}
\affil{Department of Earth and Space Science, Graduate School of Science, Osaka University, 1-1 Machikaneyama, Toyonake, Osaka 560-0043, Japan}
\author{F.~Abe}
\affil{Institute of Space-Earth Environmental Research, Nagoya University, Furo-cho, Chikusa, Nagoya, Aichi 464-8601, Japan}
\author{Y.~Asakura}
\affil{Institute of Space-Earth Environmental Research, Nagoya University, Furo-cho, Chikusa, Nagoya, Aichi 464-8601, Japan}
\author{R.~K.~Barry}
\affil{Laboratory for Exoplanets and Stellar Astrophysics, NASA/Goddard Space Flight Center, Greenbelt, MD 20771, USA}
\author{A.~Bhattacharya}
\affil{Laboratory for Exoplanets and Stellar Astrophysics, NASA/Goddard Space Flight Center, Greenbelt, MD 20771, USA}
\affil{Department of Physics, University of Notre Dame, Notre Dame, IN 46556, USA}
\author{M.~Donachie}
\affil{Department of Physics, University of Auckland, Private Bag 92019, Auckland, New Zealand}
\author{P.~Evans}
\affil{Department of Physics, University of Auckland, Private Bag 92019, Auckland, New Zealand}
\author{A.~Fukui}
\affil{Okayama Astrophysical National Astronomical Observatory, 3037-5 Honjo, Kamogata, Asakuchi, Okayama 719-0232, Japan}
\author{Y.~Hirao}
\affil{Department of Earth and Space Science, Graduate School of Science, Osaka University, 1-1 Machikaneyama, Toyonake, Osaka 560-0043, Japan}
\author{Y.~Itow}
\affil{Institute of Space-Earth Environmental Research, Nagoya University, Furo-cho, Chikusa, Nagoya, Aichi 464-8601, Japan}
\author{K.~Kawasaki}
\affil{Department of Earth and Space Science, Graduate School of Science, Osaka University, 1-1 Machikaneyama, Toyonake, Osaka 560-0043, Japan}
\author{M.~C.~A.~Li}
\affil{Department of Physics, University of Auckland, Private Bag 92019, Auckland, New Zealand}
\author{C.~H.~Ling}
\affil{Institute for Natural and Mathematical Sciences, Massey University, Private Bag 102904 North Shore Mail Centre, Auckland 0745, New Zealand}
\author{K.~Masuda}
\affil{Institute of Space-Earth Environmental Research, Nagoya University, Furo-cho, Chikusa, Nagoya, Aichi 464-8601, Japan}
\author{Y.~Matsubara}
\affil{Institute of Space-Earth Environmental Research, Nagoya University, Furo-cho, Chikusa, Nagoya, Aichi 464-8601, Japan}
\author{S.~Miyazaki}
\affil{Department of Earth and Space Science, Graduate School of Science, Osaka University, 1-1 Machikaneyama, Toyonake, Osaka 560-0043, Japan}
\author{H.~Munakata}
\affil{Institute of Space-Earth Environmental Research, Nagoya University, Furo-cho, Chikusa, Nagoya, Aichi 464-8601, Japan}
\author{Y.~Muraki}
\affil{Institute of Space-Earth Environmental Research, Nagoya University, Furo-cho, Chikusa, Nagoya, Aichi 464-8601, Japan}
\author{M.~Nagakane}
\affil{Department of Earth and Space Science, Graduate School of Science, Osaka University, 1-1 Machikaneyama, Toyonake, Osaka 560-0043, Japan}
\author{K.~Ohnishi}
\affil{Nagano National College of Technology, Nagano 381-8550, Japan}
\author{C.~Ranc}
\affil{Laboratory for Exoplanets and Stellar Astrophysics, NASA/Goddard Space Flight Center, Greenbelt, MD 20771, USA}
\author{To.~Saito}
\affil{Tokyo Metropolitan College of Industrial Technology, Tokyo 116-8523, Japan}
\author{A.~Sharan}
\affil{Department of Physics, University of Auckland, Private Bag 92019, Auckland, New Zealand}
\author{D.~J.~Sullivan}
\affil{School of Chemical and Physical Sciences, Victoria University, Wellington, New Zealand}
\author{P.~J.~Tristram}
\affil{University of Canterbury Mt John Observatory, PO Box 56, Lake Tekapo 7945, New Zealand}
\author{T.~Yamada}
\affil{Department of Physics, Faculty of Science, Kyoto Sangyo University, 603-8555 Kyoto, Japan}
\author{A.~Yonehara}
\affil{Department of Physics, Faculty of Science, Kyoto Sangyo University, 603-8555 Kyoto, Japan}
\collaboration{(MOA Collaboration)}

\begin{abstract}
    We present the result of microlensing event MOA-2016-BLG-290, which received observations from the two-wheel \emph{Kepler} (\emph{K2}), \emph{Spitzer}, as well as ground-based observatories. A joint analysis of data from \emph{K2} and the ground leads to two degenerate solutions of the lens mass and distance. This degeneracy is effectively broken once the (partial) \emph{Spitzer} light curve is included. Altogether, the lens is found to be an extremely low-mass star located in the Galactic bulge. MOA-2016-BLG-290 is the first microlensing event for which we have signals from three well-separated ($\sim1~\au$) locations. It demonstrates the power of two-satellite microlensing experiment in reducing the ambiguity of lens properties, as pointed out independently by S. Refsdal and A. Gould several decades ago.
\end{abstract}


\section{Introduction} \label{sec:introduction}

The implementation of space-based microlensing parallax has revolutionalized the field of Galactic microlensing \citep[e.g.,][]{Dong:2007,Udalski:2015a}. The same microlensing event is seen to evolve differently in views of ground-based telescopes and a space-based telescope such as \emph{Spitzer} or \emph{Kepler} because of the large separation ($\sim1~\au$) \citep{Refsdal:1966,Gould:1994,GouldHorne:2013}. This effect yields the microlensing parallax vector $\bdv{\pi_\e}$, which conveys crucial information on the lens mass and distance.

Although the \emph{Spitzer} microlensing program has been successful in terms of measuring masses of individual planetary systems \citep{Udalski:2015a,Street:2016,Shvartzvald:2017,Ryu:2017} and constraining the Galactic distribution of planets \citep{SCN:2015a,Yee:2015b,Zhu:2017}, there is a generic uncertainty in measuring $\bdv{\pi_\e}$ with a single satellite, especially in cases of single-lens events. The microlensing parallax vector $\bdv{\pi_\e}$ is directly related to the displacement between the two lens-source relative trajectories
\begin{equation}
    \bdv{\pi_\e} \approx \frac{\au}{D_\perp} \left(\frac{t_{0,\sat}-t_{0,\oplus}}{t_\e},~u_{0,\sat}-u_{0,\oplus}\right)\ ,
\end{equation}
where $D_\perp$ is the separation between the satellite and Earth perpendicular to the line of sight, $t_0$ is the time of maximum magnification, $u_0$ is the impact parameter, and subscripts ``sat'' and ``$\oplus$'' denote those seen by the satellite and Earth, respectively. Ambiguities arise because in majority cases, only $|u_0|$ (rather than $u_0$) can be constrained by the light curve, thus leading to a four-fold degeneracy in vector $\bdv{\pi_\e}$ and a two-fold degeneracy in its amplitude $\pi_\e$. Several studies have proposed ways to break these degeneracies, and others pointed out special situations in which such degeneracies do not matter (see \citealt{Yee:2015a} and references therein).

Along with proposing the idea of space-based microlensing parallax, \citet{Refsdal:1966} and \citet{Gould:1994} also pointed out that the most efficient way to break such parallax degeneracies should be to observe the same microlensing event simultaneously from another well-separated and misaligned location (satellite). The addition of a second satellite can effectively break the parallax degeneracies, especially the amplitude degeneracy.

Several decades after this idea was proposed, we finally have the chance to test it. In 2016, the two-wheel \emph{Kepler} mission (\emph{K2}, \citealt{Howell:2014}) conducted a microlensing campaign toward the Galactic bulge from April 22 to July 2, which overlapped with the \emph{Spitzer} microlensing campaign (June 18 to July 26) for nearly two weeks. With this unique opportunity, a specific program \citep{prop:2015} was developed in order to demonstrate the idea of \citet{Refsdal:1966} and \citet{Gould:1994}. In total about 30 microlensing events received observations from both satellites in addition to the dense coverage by ground-based telescopes.
\footnote{Although the two-satellite microlensing experiment with \emph{K2} and \emph{Spitzer} is the first time that we observe the same event from three well-separated locations, it is not the first time that one event was observed by ground-based and two space-based telescopes. In 2015, a few \emph{Spitzer} microlensing targets were also observed by \emph{Swift}. See \citet{Shvartzvald:2016} for the case of OGLE-2015-BLG-1319.}
This work presents the first analysis of this sample, specifically the bright single-lens event MOA-2016-BLG-290 for which the microlensing signal is detected from all three locations.
\footnote{The planetary event OGLE-2016-BLG-1190 presented in \citet{Ryu:2017} was also observed by \emph{K2}, \emph{Spitzer}, and ground-based observatories, but the \emph{K2} data did not detect the microlensing signal. Nevertheless, this non detection also led to the resolution of the parallax degeneracy. See \citet{Ryu:2017} for details.}

\section{Observations \& Data Reductions}

Microlensing event MOA-2016-BLG-290 was first identified by the MOA (Microlensing Observations in Astrophysics, \citealt{Bond:2001}) collaboration at UT 20:26 of 2016 June 1st (HJD$'=$HJD$-2450000=7541.35$), based on observations from its 1.8-m telescope with a 2.2 deg$^2$ field at Mt. John, New Zealand. About five days later, this event was also alerted as OGLE-2016-BLG-0975 by the OGLE (Optical Gravitational Lensing Experiment, \citealt{Udalski:2015b}) Collaboration through the Early Warning System \citep{Udalski:1994,Udalski:2003}, based on data taken by the 1.3-m Warsaw Telescope at the Las Campanas Observatory in Chile. 

With equatorial coordinates (R.A., decl.)$_{2000}=(18^{\rm h}04^{\rm m}57\fs01, -28\arcdeg37\arcmin40\farcs1)$ and Galactic coordinates $(l,b)_{2000}=(2\fdg40,-3\fdg50)$, this event lies inside the microlensing super stamp of the \emph{K2} Campaign 9 \citep{GouldHorne:2013,Henderson:2016}. It was therefore monitored at 30 min cadence by \emph{K2} from 2016 April 22 to May 18 (C9a, HJD$'=7501-7527$) and from May 22 to July 2 (C9b, HJD$'=7531-7572$).

Events such as MOA-2016-BLG-290 that were observed by \emph{K2} were preferentially selected during the 2016 \emph{Spitzer} microlensing campaign, for the purpose of demonstrating the two-satellite microlensing parallax concept \citep{prop:2015}. In the current case, the \emph{Spitzer} team selected it as a \emph{Spitzer} target subjectively at UT 15:03 on 2016 June 9 (HJD$'=7549.13$), following a revised protocol of \citet{Yee:2015b}. This selection turned into objective on June 12, meaning that this event met our objective selection criteria. Because of the Sun-angle limit, the \emph{Spitzer} observations were taken between HJD$'=7559.6$ and $7571.2$ at a quasi-daily cadence. All \emph{Spitzer} observations were taken in the $3.6~\micron$ channel.

The ground-based data were reduced using the standard or variant version of the image subtraction method \citep{AlardLupton:1998,Wozniak:2000,Bramich:2008}. The raw \emph{K2} light curve was extracted and modeled following the method of \citet{ZhuHuang:2017}, which is a special application of \citet{MSF:2017} and \citet{Huang:2015}. The \emph{Spitzer} data were reduced using the software that was customized for the microlensing program \citep{SCN:2015b}.

\section{Breaking Parallax Degeneracy with Two-Satellite Experiment}

\begin{deluxetable}{lDDDD}
\tabletypesize{\scriptsize}
\tablecaption{Best-fit parameters of microlensing models. With only ground-based and \emph{K2} data, all four solutions are allowed, but only the first two [$(+,+)$ and $(-,-)$] survive once \emph{Spitzer} data and the the associated source $I-[3.6\micron]$ color constraint are taken into account.
\label{tab:fourfold}}
\tablehead{
\colhead{Parameters} & \multicolumn2c{$(+,+)$} & \multicolumn2c{$(-,-)$} & \multicolumn2c{$(+,-)$} & \multicolumn2c{$(-,+)$}
}
\decimals
\startdata
$t_{0,\oplus}-7552$ & $0.380(4)$ & $0.378(4)$ & $0.375(4)$ & $0.383(4)$ \\
$u_{0,\oplus}$ & $0.7032(6)$ & $-0.7033(6)$ & $0.7036(4)$ & $-0.7031(6)$ \\
$t_\e$ & $6.370(8)$ & $6.370(9)$ & $6.379(9)$ & $6.373(9)$ \\
$\pien$ & $0.180(5)$ & $-0.200(5)$ & $-2.199(6)$ & $2.156(4)$ \\
$\piee$ & $-0.150(4)$ & $-0.131(4)$ & $-0.139(4)$ & $-0.370(4)$ \\
$I-[3.6\micron]$ & $-5.29(13)$ & $-5.35(13)$ & $-4.89(13)$ & $-4.53(13)$ \\
\enddata
\end{deluxetable}

\begin{figure*}
\epsscale{0.9}
\plotone{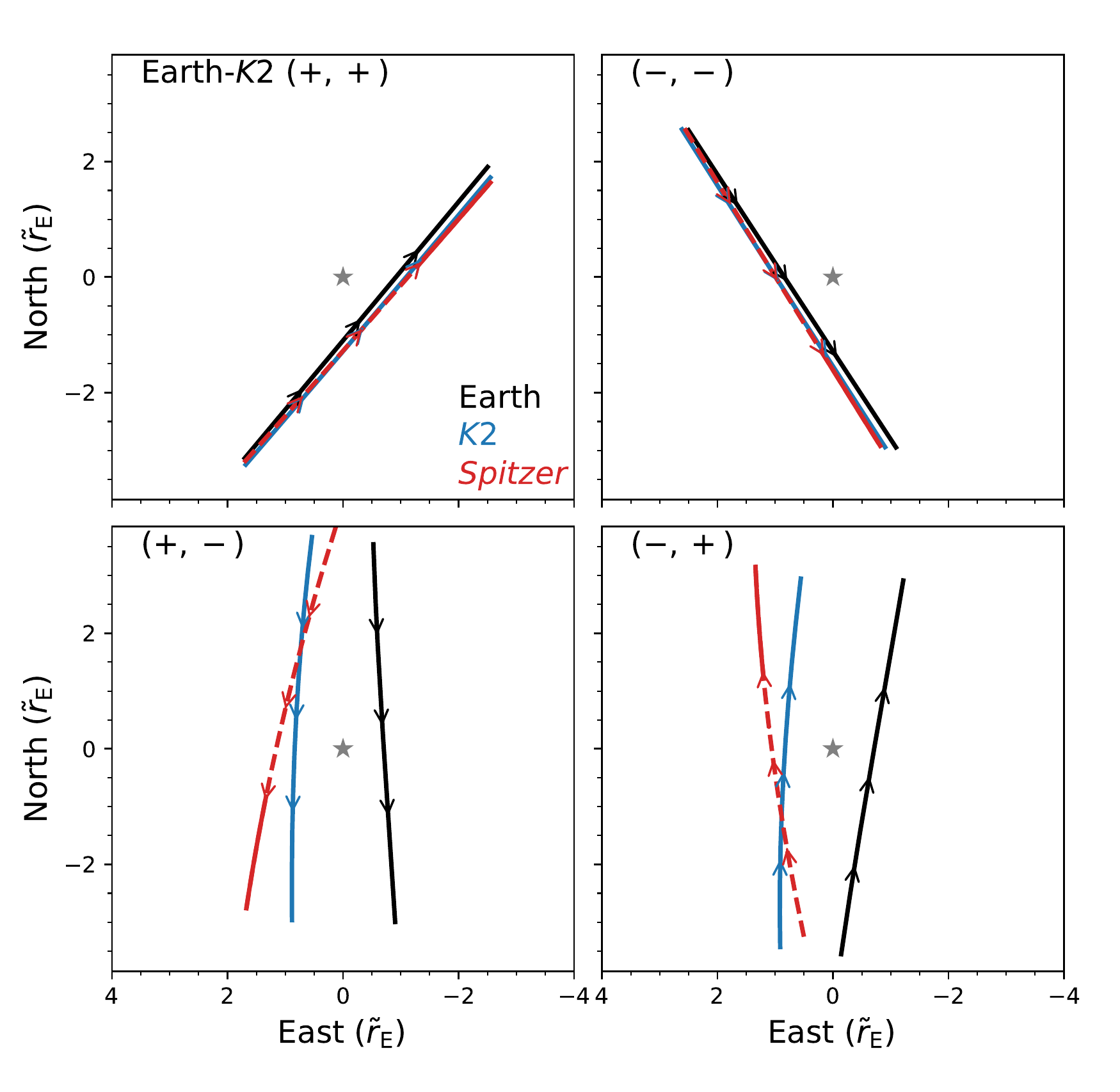}
\caption{Trajectories of three observers (Earth, \emph{K2}, and \emph{Spitzer}) with respect to the aligned lens and source position (marked as asterisks). These are the \textit{geocentric} views of microlensing geometries \citep{Gould:2004}. The four solutions allowed by the ground-based and \emph{K2} data are shown individually, and the predicted \emph{Spitzer} positions are shown in red, with the solid line denoting the time span of actual \emph{Spitzer} observations. For each trajectory, there are three arrows indicating the direction of motion at three different epochs: HJD$'=7540$, 7550, and 7560, respectively. The trajectories are oriented so that north is up and east is left (see \citealt{Skowron:2011} for the sign convention of $u_0$). We note that the Earth-\emph{K2}-\emph{Spitzer} relative positions (in AU) are the same in all plots, and that they are simply scaled differently in all plots for the given parallax measurements.
\label{fig:geometry-geo}}
\end{figure*}

\begin{figure*}
\epsscale{0.9}
\plotone{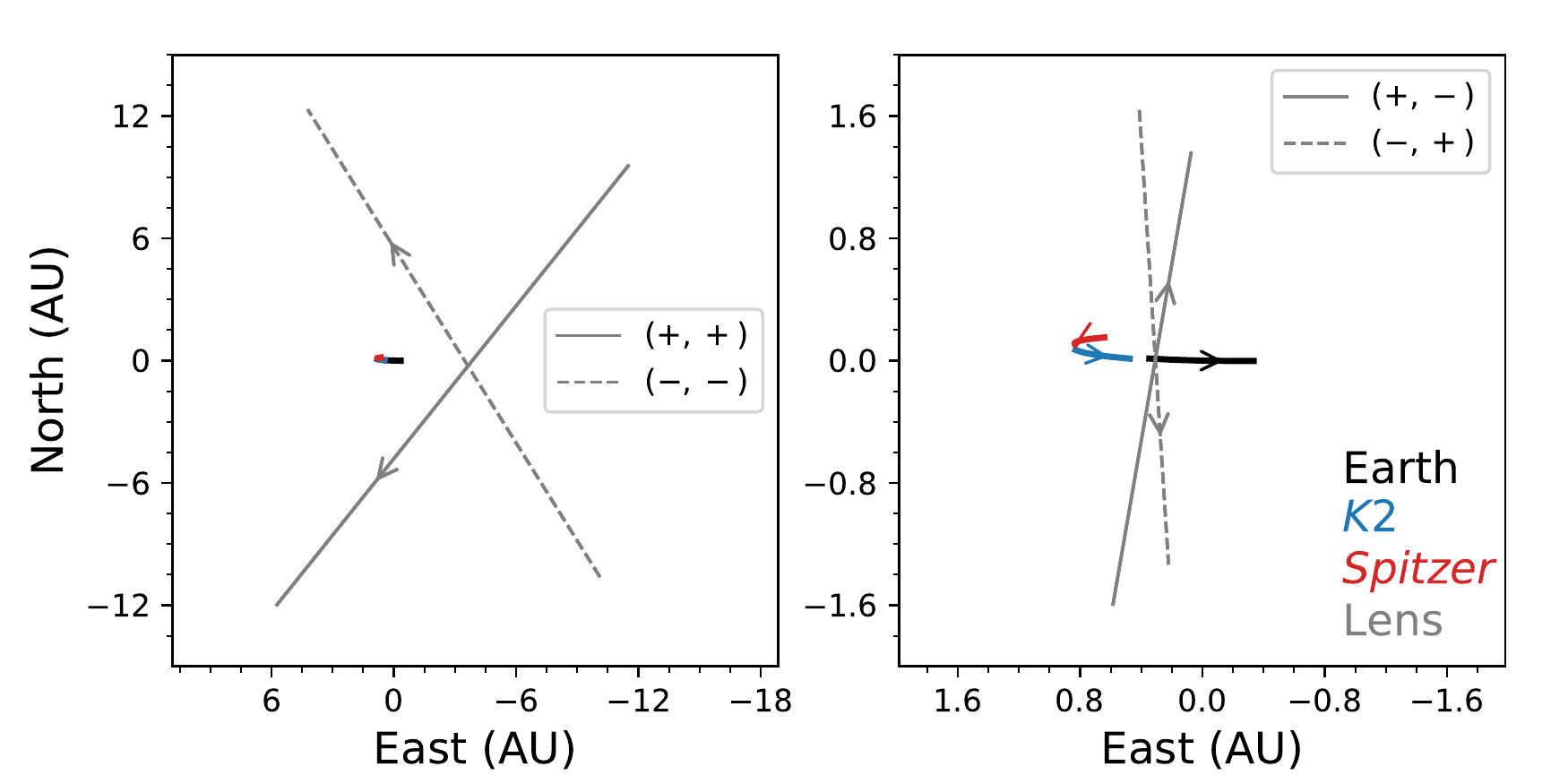}
\caption{Trajectories of the lens with respect to the aligned source and observer (Earth at $t_{0,\oplus}$), and the motions of all three observers relative to the same reference point. These are the \textit{heliocentric} views of microlensing geometries \citep{SCN:2016}. Again, north is up and east is left, but we use the same physical scale for all solutions so that the motions of observers are the same.
For each curve, the arrow indicates the position of the object as well as its directory of motion. Now the Earth-trailing orbits of \emph{K2} and \emph{Spitzer} are clearly seen.
\label{fig:geometry-hel}}
\end{figure*}

First, as a proof of concept of the two-satellite microlensing parallax method, we choose to only model the ground-based and \emph{K2} data, and then compare the predicted \emph{Spitzer} light curve with the actual \emph{Spitzer} data.

The modeling of ground-based and \emph{K2} data follows the methodology of \citet{ZhuHuang:2017}, but with a minor modification. According to the OGLE-III Catalog of Variable Stars \citep{Soszynski:2013}, a low-amplitude (0.034~mag) long-period (342.5 days) variable, OGLE-BLG-LPV-202211, sits only $10\arcsec$ (or 2.5 \emph{K2} pixels) away from the location of MOA-2016-BLG-290. Due to the broad point spread function and the unstable pointing of the \emph{K2} spacecraft, this variable star affects the raw \emph{K2} light curve that we extracted. To minimize the influence of this nearby variable star, we introduce an additional term that scales linearly with time into the model of \emph{K2} raw light curve (Equation~1 of \citealt{ZhuHuang:2017}). Following \citet{ZhuHuang:2017}, we also include a constraint on the source $K_p-I$ color, which is derived from the source $V-I$ color from OGLE photometry.

With only data from the ground and from \emph{K2} included in the modeling, the four-fold degeneracy emerges. These are generally denoted as (Earth-$K2$) $(\pm,\pm)$ solutions, with the first sign and the second sign indicating the sign of $u_{0,\oplus}$ and $u_{0,K2}$ in the geocentric frame, respectively \citep{Zhu:2015}. The microlensing parameters for these solutions are given in Table~\ref{tab:fourfold}, and the microlensing geometries are shown in Figures~\ref{fig:geometry-geo} and \ref{fig:geometry-hel} for different choices of reference frame. As these plots illustrate, the four solutions are distinct in terms of the velocity vector of the lens-source relative motion, which is directly determined by the parallax vector $\bdv{\pie}$. If only the amplitude of parallax is considered, the four-fold degeneracy essentially collapses to two-fold \citep{Gould:1994}, which we denote as ``small $\pie$'' [$(+,+)$ and $(-,-)$] and ``large $\pie$'' [$(+,-)$ and $(-,+)$].

\begin{figure*}
\epsscale{1.1}
\plottwo{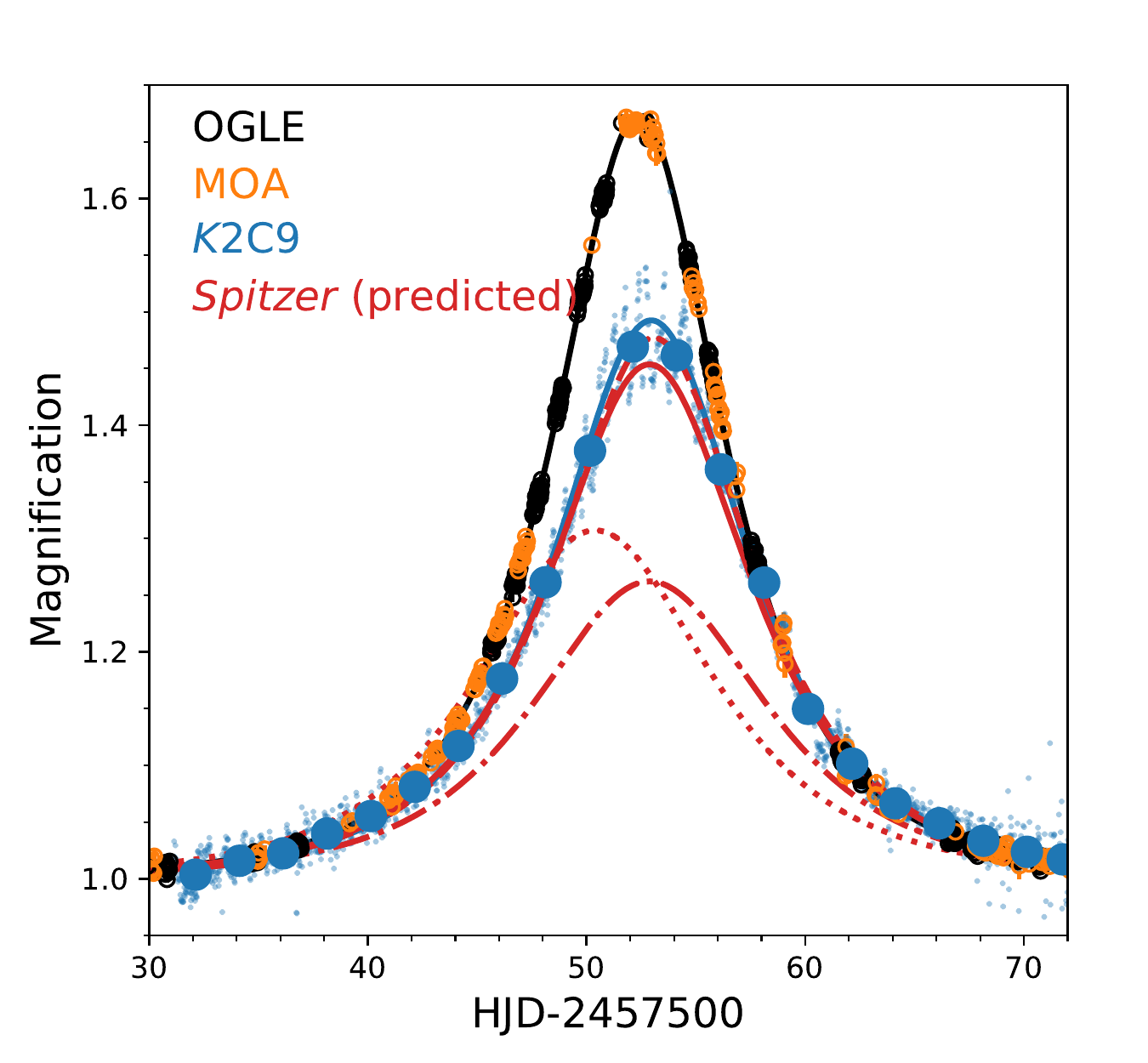}{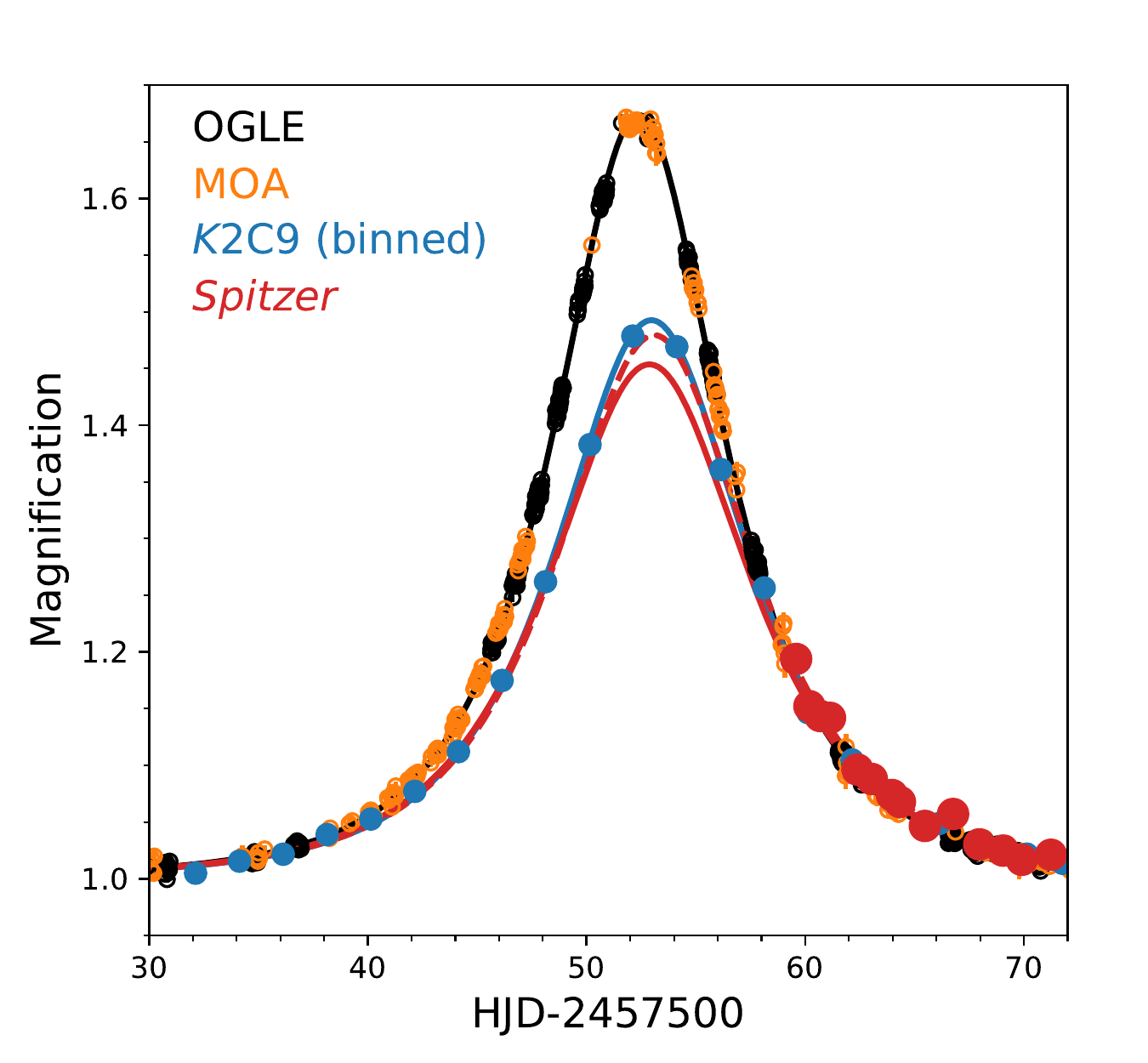}
\caption{\textit{Left panel:} predicted \emph{Spitzer} light curves (in red) based on the modeling of the ground-based (OGLE in black and MOA in orange) and \emph{K2}C9 data (in blue). The raw \emph{K2} data are shown as blue open dots, and the binned data are shown as blue solid dots with error bars. The four predicted \emph{Spitzer} light curves are plotted with different line styles: solid, dashed, dash-dot, and dotted for Earth-\emph{K2} $(+,+)$, $(-,-)$, $(+,-)$, and $(-,+)$ solutions, respectively.
\textit{Right panel:} Data and the best-fit models for all observatories. Here we only show the re-binned \emph{K2}C9 data, and the remaining labels are the same as those in the left panel. Once all data and the color constraints are included, only the ``small $\pie$'' solutions, $(+,+)$ and $(-,-)$, are allowed.
\label{fig:lcs}}
\end{figure*}

Because \emph{Spitzer}'s position relative to Earth and \emph{K2} is well known, we can then ``predict'' the microlensing light curve that \emph{Spitzer} would see for all four solutions. These predicted \emph{Spitzer} light curves are shown in the left panel of Figure~\ref{fig:lcs} together with the ground-based (OGLE \& MOA) and \emph{K2} light curves. Given the different behaviors of the predicted light curves, \emph{Spitzer} observations would in principle pick out the correct solution if this third observer had a full coverage of the event light curve. Unfortunately, \emph{Spitzer} was not able to observe this event until HJD$'=7559.5$ because of its Sun-angle limit, and therefore it only captured the falling tail of the light curve. Such a partial light curve can be fit by all four solutions equally well, if no other information is provided.

Fortunately, with the known properties of the source star, we are able to at least break the degeneracy in the parallax amplitude $\pie$ (i.e., ``small $\pie$'' vs. ``large $\pie$''). The \emph{Spitzer} (as well as OGLE and MOA) flux is modeled by
\begin{equation}
F(t) = F_\s \cdot A(t) + F_\b\ .
\end{equation}
Here $F_\s$ is the source flux in given observatory/bandpass, $F_\b$ is the flux that is within the aperture but unrelevant to the microlensing effect, and $A(t)$ is the microlensing magnification at given time $t$. For the same set of \emph{Spitzer} measurements, a different magnification behavior would suggest a different source brightness $F_\s$ (and so source color $I-[3.6\micron]$, since the source $I$ magnitude is well determined). With the predicted \emph{Spitzer} light curves from the previous step, the source $I-[3.6\micron]$ colors are then estimated for all four solutions, and they are listed in Table~\ref{tab:fourfold} as well. The uncertainty on the source color is dominated by uncertainties on the \emph{Spitzer} observations. For the two groups of solutions, the inferred source $I-[3.6\micron]$ colors are statistically different at $>3~\sigma$ level.

\begin{figure}
\epsscale{1.2}
\plotone{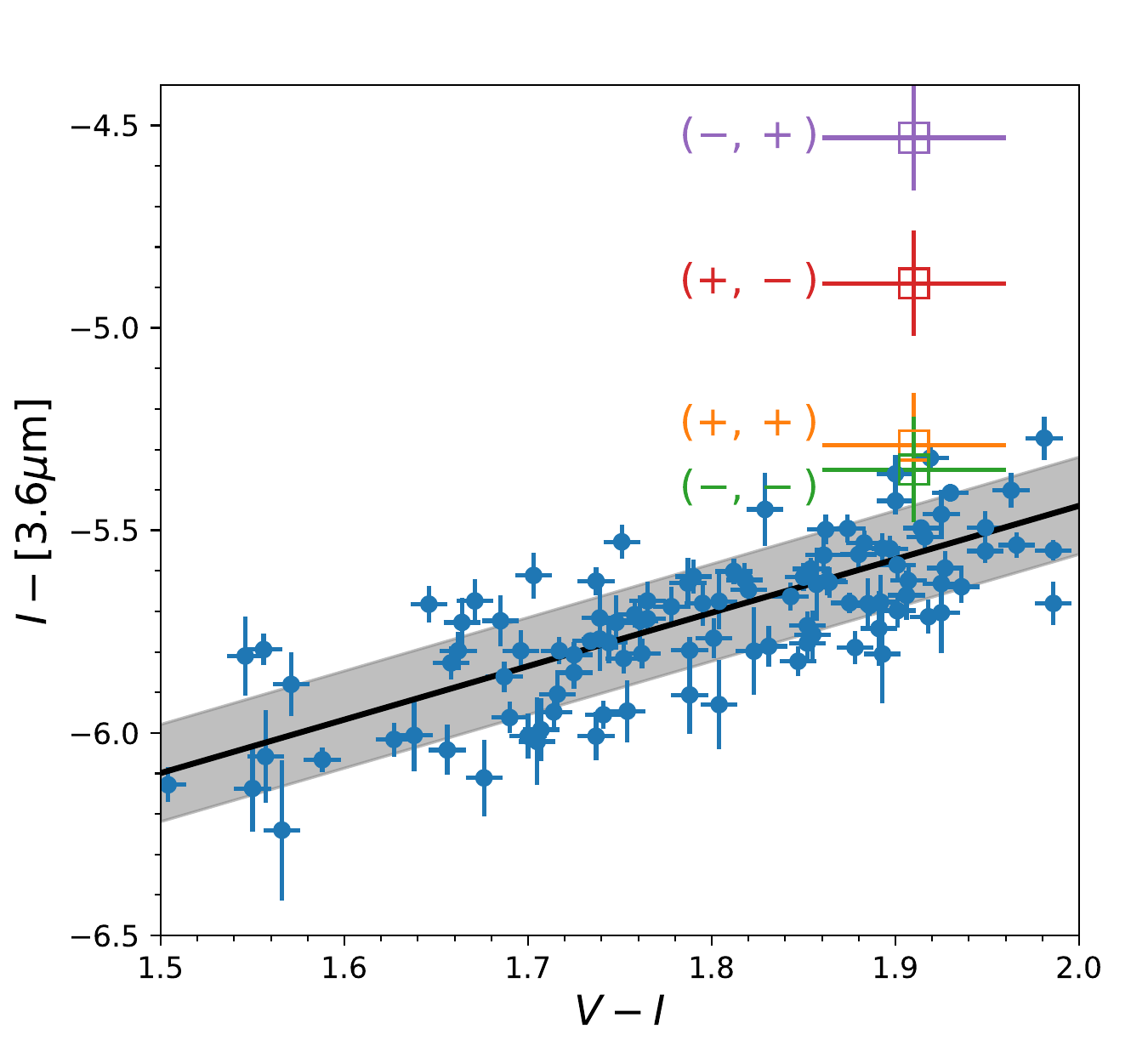}
\caption{This figure shows the stellar $I-[3.6\micron]$ vs. $V-I$ color-color relation as well as the data points used to derive it. The shaded region remarks the 1-$\sigma$ uncertainty of this color-color relation. The open squares are the source colors inferred from the four solutions. The colors of the ``small $\pie$'' solutions $(++ \& --)$ are consistent with this independent color measurement, while those of the ``large $\pie$'' solutions $(+- \& -+)$ are inconsistent at $>3\sigma$ level.
\label{fig:colors}}
\end{figure}

We then derive the source color 
\begin{equation}
I-[3.6\micron]=-5.56\pm0.12
\end{equation}
from a model-independent way, by substituting the source $V-I$ color into the stellar $I-[3.6\micron]$ vs. $V-I$ color-color relation. This relation is established based on neighboring field stars with similar properties \citep[see details in][]{SCN:2015b}. The deviations between this color measurement and inferred colors are $1.5\sigma$, $1.2\sigma$, $3.8\sigma$, and $5.8\sigma$ for $(+,+)$, $(-,-)$, $(+,-)$, and $(-,+)$ solutions, respectively. Therefore, the ``large $\pie$'' solutions can be securely rejected, and only the ``small $\pie$'' solutions are allowed. See Figure~\ref{fig:colors} for the illustration of the color determination and comparison.

As a final step, we model data from all observatories (OGLE, MOA, \emph{K2} and \emph{Spitzer}) simultaneously. The microlensing parameters are almost identical to those for Earth-\emph{K2} $(+,+)$ and $(-,-)$ solutions, and therefore are not listed separately here. The best-fit models and all data sets are illustrated in the right panel of Figure~\ref{fig:lcs}.

\section{Discussion}

In this work, we present the analysis of the first microlensing event that has detected signals from at least three well-separated ($\sim1~\au$) locations, which in the current case are Earth, \emph{K2}, and \emph{Spitzer}. With data from the third well-separated observer, we have demonstrated that the generic parallax degeneracy arising in the single-satellite microlensing parallax experiment is effectively broken. This is essentially the first realization of the decades-old idea proposed by \citet{Refsdal:1966} and \citet{Gould:1994} independently.

For the current event, we could only break the degeneracy in the (two-fold) parallax amplitude rather than the degeneracy in the (four-fold) parallax vector. This is partly because this event could not be observed by \emph{Spitzer} until it was almost finished, but mostly because the event is near the ecliptic plane and the Earth-\emph{K2}-\emph{Spitzer} configuration is nearly colinear \citep{GaudiGould:1997}. Nevertheless, it is the amplitude of the microlensing parallax that matters in determining the lens properties, and therefore being able to break the degeneracy in parallax amplitude has already enabled a better determination of the lens properties, such as lens mass. We use the current case as an example. Using the Galactic model and the Bayesian method of \citet{Zhu:2017}, we estimate the lens of MOA-2016-BLG-290 has a mass $M_\lens=77_{-23}^{+34}~M_\jup$ and distance $D_\lens=6.8\pm0.4~$kpc for accepted solutions. These values correspond to a brown dwarf or extremely low-mass star likely in the Galactic bulge. However, the other two solutions, had they been correct, would suggest a high-mass ($7_{-3}^{+4}~M_\jup$) planet in the near disk ($D_\lens=2.5\pm0.8~$kpc). The ability to break the $\pie$ amplitude degeneracy really reduces the uncertainties on the lens properties.

Among other proposed methods \citep{Gould:1995,Gould:1999,GouldYee:2012,Yee:2015a,SCN:2015a}, obtaining observations from a third location has always been considered the most efficient in breaking the parallax degeneracy in satellite microlensing experiments. It is nevertheless difficult to do so for a large number of events because of obvious economical reasons. In the absence of this method, the Rich argument can be applied for statistical purposes \citep{SCN:2015a,Zhu:2017}. In fact, for the present case, the application of the Rich argument would also argue for an extremely low probability ($1\%)$ of the ``large $\pie$'' solutions. However, the resolution of degeneracy is important whenever precise knowledge is required of an individual event. Therefore, the $\sim$30 events observed in the 2016 two-satellite microlensing experiment are very valuable. This ensemble can be used to refine the microlensing parallax method, which is the foundation of the on-going \emph{Spitzer} microlensing project and will likely be a crucial component of the future space-based microlensing surveys.

Regardless, one may wonder when will be the next time to apply this two-satellite parallax method? Within the predictable period, one could only hope to apply this method one decade from now to the Wide Field InfaRed Survey Telescope (WFIRST, \citealt{Spergel:2015}) and Euclid \citep{Penny:2013}, although the situation will be largely different because these telescopes will likely be at Earth-Sun L2 point (i.e., a much shorter baseline, see \citealt{ZhuGould:2016} for a detailed analysis of WFIRST parallax. See also \citealt{Yee:2013}). Nevertheless, the history of microlensing parallax has already proved that fantastic scientific ideas will never be buried.

\acknowledgements
Work by W.Z. and A.G. were supported by NSF grant AST-1516842. 
Work by S.C.N. and A.G. were supported by JPL grant 1500811.
R.P. acknowledges support from K2 Guest Observer program under NASA grant NNX17AF72G.
Work by Y.S. was supported by an appointment to the NASA Postdoctoral Program at the Jet Propulsion Laboratory, California Institute of Technology, administered by Universities Space Research Association through a contract with NASA.
Work by C.R. was supported by an appointment to the NASA Postdoctoral Program at the Goddard Space Flight Center, administered by USRA through a contract with NASA.
This paper includes data collected by the Kepler mission. Funding for the Kepler mission is provided by the NASA Science Mission directorate. Some of the data presented in this paper were obtained from the Mikulski Archive for Space Telescopes (MAST). STScI is operated by the Association of Universities for Research in Astronomy, Inc., under NASA contract NAS5-26555. Support for MAST for non-HST data is provided by the NASA Office of Space Science via grant NNX09AF08G and by other grants and contracts.
OGLE project has received funding from the National Science Centre, Poland, grant MAESTRO 2014/14/A/ST9/00121 to A.U..
The MOA project is supported by JSPS Kakenhi grants JP24253004, JP26247023, JP16H06287, JP23340064 and JP15H00781 and by the Royal Society of New Zealand Marsden Grant MAU1104.

\end{CJK*}

\begin{thebibliography}{}
\bibitem[Alard \& Lupton(1998)]{AlardLupton:1998} Alard, C., \& Lupton, R.~H.\ 1998, \apj, 503, 325 
\bibitem[Bond et al.(2001)]{Bond:2001} Bond, I.~A., Abe, F., Dodd, R.~J., et al.\ 2001, \mnras, 327, 868 
\bibitem[Bramich(2008)]{Bramich:2008} Bramich, D.~M.\ 2008, \mnras, 386, L77 
\bibitem[Calchi Novati et al.(2015a)]{SCN:2015a} Calchi Novati, S., Gould, A., Udalski, A., et al.\ 2015, \apj, 804, 20 
\bibitem[Calchi Novati et al.(2015b)]{SCN:2015b} Calchi Novati, S., Gould, A., Yee, J.~C., et al.\ 2015, \apj, 814, 92 
\bibitem[Calchi Novati \& Scarpetta(2016)]{SCN:2016} Calchi Novati, S., \& Scarpetta, G.\ 2016, \apj, 824, 109 
\bibitem[Dong et al.(2007)]{Dong:2007} Dong, S., Udalski, A., Gould, A., et al.\ 2007, \apj, 664, 862 
\bibitem[Gaudi \& Gould(1997)]{GaudiGould:1997} Gaudi, B.~S., \& Gould, A.\ 1997, \apj, 477, 152 
\bibitem[Gould(1994)]{Gould:1994} Gould, A.\ 1994, \apjl, 421, L75 
\bibitem[Gould(1995)]{Gould:1995} Gould, A.\ 1995, \apjl, 441, L21 
\bibitem[Gould(1999)]{Gould:1999} Gould, A.\ 1999, \apj, 514, 869 
\bibitem[Gould(2004)]{Gould:2004} Gould, A.\ 2004, \apj, 606, 319 
\bibitem[Gould \& Yee(2012)]{GouldYee:2012} Gould, A., \& Yee, J.~C.\ 2012, \apjl, 755, L17 
\bibitem[Gould \& Horne(2013)]{GouldHorne:2013} Gould, A., \& Horne, K.\ 2013, \apjl, 779, L28 
\bibitem[Gould et al.(2015)]{prop:2015} Gould, A., Yee, J., \& Carey, S.\ 2015, Spitzer Proposal, 12015 
\bibitem[Henderson et al.(2016)]{Henderson:2016} Henderson, C.~B., Poleski, R., Penny, M., et al.\ 2016, \pasp, 128, 124401 
\bibitem[Howell et al.(2014)]{Howell:2014} Howell, S.~B., Sobeck, C., Haas, M., et al.\ 2014, \pasp, 126, 398 
\bibitem[Huang et al.(2015)]{Huang:2015} Huang, C.~X., Penev, K., Hartman, J.~D., et al.\ 2015, \mnras, 454, 4159 
\bibitem[Penny et al.(2013)]{Penny:2013} Penny, M.~T., Kerins, E., Rattenbury, N., et al.\ 2013, \mnras, 434, 2 
\bibitem[Refsdal(1966)]{Refsdal:1966} Refsdal, S.\ 1966, \mnras, 134, 315 
\bibitem[Ryu et al.(2017)]{Ryu:2017} Ryu, Y.-H., et al.\ 2017\ ApJ submitted
\bibitem[Shvartzvald et al.(2016)]{Shvartzvald:2016} Shvartzvald, Y., Li, Z., Udalski, A., et al.\ 2016, \apj, 831, 183 
\bibitem[Shvartzvald et al.(2017)]{Shvartzvald:2017} Shvartzvald, Y., Yee, J.~C., Calchi Novati, S., et al.\ 2017, \apjl, 840, L3 
\bibitem[Skowron et al.(2011)]{Skowron:2011} Skowron, J., Udalski, A., Gould, A., et al.\ 2011, \apj, 738, 87 
\bibitem[Soares-Furtado et al.(2017)]{MSF:2017} Soares-Furtado, M., Hartman, J.~D., Bakos, G.~{\'A}., et al.\ 2017, \pasp, 129, 044501 
\bibitem[Soszy{\'n}ski et al.(2013)]{Soszynski:2013} Soszy{\'n}ski, I., Udalski, A., Szyma{\'n}ski, M.~K., et al.\ 2013, \actaa, 63, 21 
\bibitem[Spergel et al.(2015)]{Spergel:2015} Spergel, D., Gehrels, N., Baltay, C., et al.\ 2015, arXiv:1503.03757 
\bibitem[Street et al.(2016)]{Street:2016} Street, R.~A., Udalski, A., Calchi Novati, S., et al.\ 2016, \apj, 819, 93 
\bibitem[Udalski et al.(1994)]{Udalski:1994} Udalski, A., Szymanski, M., Kaluzny, J., et al.\ 1994, \actaa, 44, 227 
\bibitem[Udalski(2003)]{Udalski:2003} Udalski, A.\ 2003, \actaa, 53, 291
\bibitem[Udalski et al.(2015a)]{Udalski:2015a} Udalski, A., Yee, J.~C., Gould, A., et al.\ 2015, \apj, 799, 237 
\bibitem[Udalski et al.(2015b)]{Udalski:2015b} Udalski, A., Szyma{\'n}ski, M.~K., \& Szyma{\'n}ski, G.\ 2015, \actaa, 65, 1 
\bibitem[Wozniak(2000)]{Wozniak:2000} Wozniak, P.~R.\ 2000, \actaa, 50, 421 
\bibitem[Yee(2013)]{Yee:2013} Yee, J.~C.\ 2013, \apjl, 770, L31 
\bibitem[Yee et al.(2015a)]{Yee:2015a} Yee, J.~C., Udalski, A., Calchi Novati, S., et al.\ 2015, \apj, 802, 76 
\bibitem[Yee et al.(2015b)]{Yee:2015b} Yee, J.~C., Gould, A., Beichman, C., et al.\ 2015, \apj, 810, 155 
\bibitem[Zhu \& Gould(2016)]{ZhuGould:2016} Zhu, W., \& Gould, A.\ 2016, Journal of Korean Astronomical Society, 49, 93 
\bibitem[Zhu et al.(2015)]{Zhu:2015} Zhu, W., Udalski, A., Gould, A., et al.\ 2015, \apj, 805, 8 
\bibitem[Zhu et al.(2017a)]{Zhu:2017} Zhu, W., Udalski, A., Calchi Novati, S., et al.\ 2017, arXiv:1701.05191 
\bibitem[Zhu et al.(2017b)]{ZhuHuang:2017} Zhu, W., Huang, C.~X., Udalski, A., et al.\ 2017, \pasp, 129, 104501 
\end{thebibliography}
\end{document}